\documentclass{PoS}
\usepackage{amsmath}
\usepackage{slashed}

\newcommand\br[1]{\mathopen{\left\langle#1\right|}}
\newcommand\ket[1]{\mathclose{\left|#1\right\rangle}}

\newcommand\url[1]{\href{#1}{#1}}

\newcommand\slurp[1]{#1}
\newcommand\sslurp[2]{#1{#2}}
{\catcode`/=\active \expandafter}%
\slurp{\newcommand/}{/\penalty1000\hskip0pt\relax}
{\catcode`:=\active \expandafter}%
\slurp{\newcommand:}{:\penalty1000\hskip0pt\relax}

\newcommand\addspace{\ifcat\nextchar a\spacefactor999. \else.\fi}
{\catcode`\.=\active \expandafter}%
\slurp{\newcommand.}{\futurelet\nextchar\addspace}

\newcommand\myslash{/} \newcommand\mycolon{:}
\newcommand\doi{{\catcode`/=\active \catcode`:=\active \expandafter}\sslurp\realdoi}
{\catcode`/=\active \catcode`:=\active \expandafter}%
\slurp{\newcommand\realdoi[1]{{\let/=\myslash \let:=\mycolon
                               \edef\raw{{http://dx.doi.org/#1}}\expandafter}%
                               \expandafter\href\raw{doi:#1}}}
\newcommand\eprint[2]{{\escapechar-1%
                       \edef\a{\expandafter\string\csname arXiv\endcsname}%
                       \edef\b{\expandafter\string\csname #1\endcsname}%
                       \edef\c{\expandafter\string\csname #2\endcsname}%
                       \edef\d{\noexpand\href{http://arXiv.org/abs/\c}}%
                       \ifx\a\b\expandafter\d\fi{\tt #1:#2}}}
\newcommand\linkdoi[2]{\href{http://dx.doi.org/#1}{#2}}
\def\linkurl#{{\catcode`\%=12\catcode`\_=12\relax\expandafter}\readurl}
\newcommand\readurl[1]{\href{#1}}

\let\ifpdf=\undefined
\makeatletter\let\orig@footnotemark\@footnotemark
\usepackage{hyperref}
\makeatletter\let\saved@footnotemark\@footnotemark
\AtBeginDocument{\let\@footnotemark\saved@footnotemark}
\let\@footnotemark\orig@footnotemark
\let\savedhyper@@link\hyper@@link
\renewcommand\hyper@@link{\forever\savedhyper@@link}
\newcommand\preserve{\noexpand\forever\noexpand}
\newcommand\notrelax{\noexpand\notrelax}
\newcommand\forever{{\let\notrelax\relax\expandafter}\ifx\notrelax\relax\else\expandafter\preserve\fi}
\renewcommand\intPlink[2]{{#2}}

\title{Theoretical Bounds on New Four-Fermion Interactions and TeV Scale Physics}

\ShortTitle{Theoretical Bounds on New Four-Fermion Interactions and TeV Scale Physics}

\author{\speaker{Tanmoy Bhattacharya}, Rajan Gupta and Anosh Joseph\\
        Theoretical Division, Los Alamos National Laboratory, Los Alamos, NM 87545}

\author{Huey-Wen Lin and Saul D. Cohen\\
        Department of Physics, University of Washington, Seattle, WA 98195-1560}

\abstract{The standard model weak interactions can be described by
  four-fermion \(V-A\) operators at low energies. New physics at the
  TeV scale can, however, generate the other Lorentz structures. In
  this talk, we review the constraints on such interactions from
  nuclear and hadronic decays, as well as from collider searches.
  Currently the most stringent bounds come from the analysis of the
  \(0^+\to0^+\) nuclear and the \(\pi\to e\nu\gamma\) radiative pion
  decays. In the near future, the ultracold neutron beta decay
  experiments and the direct LHC measurements will compete in
  setting the most stringent bounds, provided, however, that the
  neutron-to-proton non-perturbative transition matrix elements can be
  calculated to a level of 10--20\% accuracy.}

\FullConference{XXIX International Symposium on Lattice Field Theory \\
   July 10--16 2011\\
    Squaw Valley, Lake Tahoe, California}

\begin{document}

\section{Effective Lagrangian for the Charge-Current Interactions}
We follow the notation of Ref.~\cite{Cirigliano:2009wk}, which
identified a minimal basis for the \(SU(2)_L\times U(1)_Y\)-invariant
dimension-six operators contributing to low-energy charged-current
processes. In particular, we study only theories that do not violate
\(CP\) and conserve baryon and lepton numbers at this level, and that
do not contain light right-handed neutrinos.  In such theories, we can
write the part of this charged-current Lagrangian coupling
quarks to leptons as
\begin{eqnarray}
{\cal L}_{\rm CC} &=&  \frac{-g^2}{2 M_W^2} V_{ij} \Bigg[
 \Big(1 + [v_L]_{\ell \ell' ij} \Big) \bar{\ell}_L \gamma_\mu
 \nu_{\ell' L}  \bar{u}_L^i \gamma^\mu d_L^j
  +  [v_R]_{\ell \ell' ij}  \bar{\ell}_L \gamma_\mu  \nu_{\ell'
   L} \bar{u}_R^i \gamma^\mu d_R^j
  \nonumber\\
&+&  [s_L]_{\ell \ell' ij}  \bar{\ell}_R   \nu_{\ell' L}
  \bar{u}_R^i  d_L^j
  + [s_R]_{\ell \ell' ij}  \bar{\ell}_R   \nu_{\ell' L}
  \bar{u}_L^i  d_R^j
   \nonumber \\
&+&   [t_L]_{\ell \ell' ij}  \bar{\ell}_R   \sigma_{\mu \nu}
  \nu_{\ell' L}    \bar{u}_R^i   \sigma^{\mu \nu} d_L^j
\Bigg] + {\rm h.c.}\,,
\end{eqnarray}
where we have suppressed the color indices and used the notation
\(\sigma^{\mu\nu} = i[\gamma^\mu,\gamma^\nu]/2\). Further, \(g\) is
the weak coupling, \(M_W\) is the mass of the \(W\)-boson, \(V_{ij}\)
refer to the CKM matrix elements, \(L\) and \(R\) to the chiral
projections, \(\ell\) and \(\ell'\) to the lepton families, \(i\) and
\(j\) to the quark families, \(u\) and \(d\) to the generic up and
down type quarks, and \(\ell\) and \(\nu_\ell\) to the charged leptons
and neutrinos, respectively. This effective theory contains five
families of effective couplings: \(v_L\), \(v_R\), \(s_L\), \(s_R\),
and \(t_L\), which are expected to be of order
\(v^2/\Lambda^2 \sim 10^{-3}\), where \(v\) is the Higgs VEV and
\(\Lambda\) is the scale of new physics.  Even though we write only
the charged current sector here, we note that due to the \(SU(2)\)
invariance of the interactions, the same effective couplings also
mediate neutral current interactions that can be used to constrain
them. 

For the most part we will be interested only in the first family of
the quarks and work to linear order in the effective BSM couplings.
Also suppressing the lepton family indices, we can write
\begin{eqnarray}
{\cal L}_{\rm CC} &=& - \frac{G_F^{(0)} V_{ud}}{\sqrt{2}} \Big(1 + \epsilon_V  \Big)
\Bigg[ \bar{\ell}  \gamma_\mu  (1 - \gamma_5)   \nu
       \cdot \bar{u}   \Big[ \gamma^\mu (1 - \gamma_5) + \big(\epsilon_V-\epsilon_A \big)
       \gamma^\mu \gamma_5 \Big] d \nonumber\\
     &+& \bar{\ell}  (1 - \gamma_5) \nu_{\ell}
        \cdot \bar{u}  \Big[  \epsilon_S  -   \epsilon_P \gamma_5 \Big]  d
+ \epsilon_T \bar{\ell}   \sigma_{\mu \nu} (1 - \gamma_5)
\nu    \cdot  \bar{u}   \sigma^{\mu \nu} (1 - \gamma_5) d
\Bigg]+{\rm h.c.}\,,
\end{eqnarray}
where \(G^{(0)}_F\) is the tree-level Fermi constant, \(\epsilon_{V,A} \equiv
v_L \pm v_R\), \(\epsilon_{S,P} \equiv s_L \pm s_R\),  \(\epsilon_T
\equiv t_L\).  In this notation, \(\epsilon_V\) affects the overall
normalization of the Fermi constant and is constrained both from
low-energy and Z-pole observables. The right handed vector coupling,
\(\epsilon_V - \epsilon_A\), however, only affects the ratio of
Axial-to-Vector couplings and constraining it meaningfully from
hadronic physics needs determination of the rato of vector and axial
charges to better than \(10^{-3}\) level.  The rest of the couplings
\(\epsilon_{S,P,T}\) violate chirality and, hence, their intereference
with the Standard Model interactions is suppressed by \(m_\ell/E\);
consequently, they are suppressed in high-energy experiments, but
remain accessible in pion decays and asymmetry measurements in beta
decays. 

\section{Collider Limits}
\begin{figure}[t]
\hfill
\begin{minipage}{0.45\hsize}
\begin{center}
\includegraphics[height=0.2\vsize]{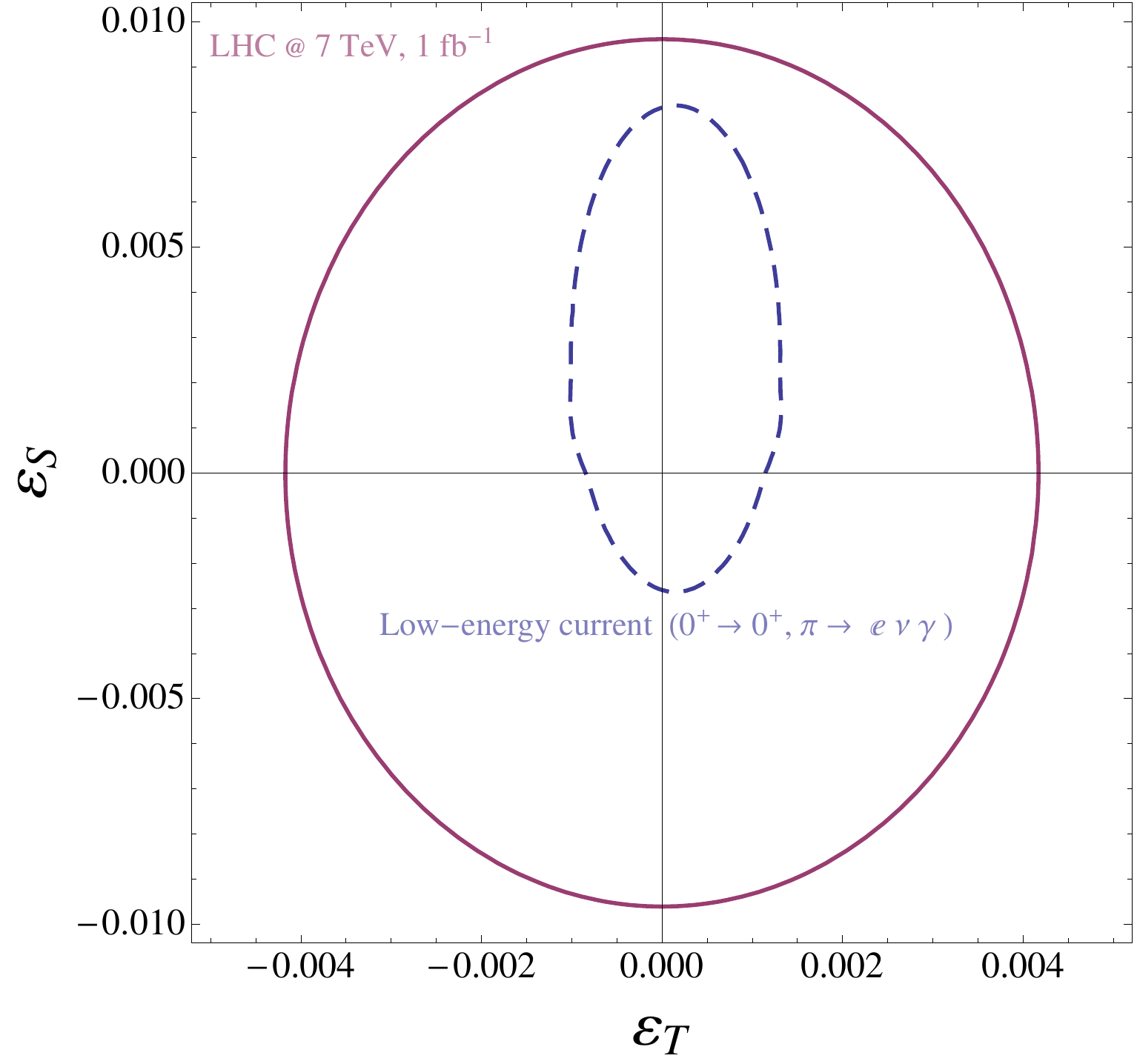}
\end{center}
\caption{The bounds on the BSM scalar and tensor interactions obtained
  at LHC compared to those from low-energy measurements.}
\label{BSMlow}
\end{minipage}\hfill
\begin{minipage}{0.45\hsize}
\begin{center}
\includegraphics[height=0.2\vsize]{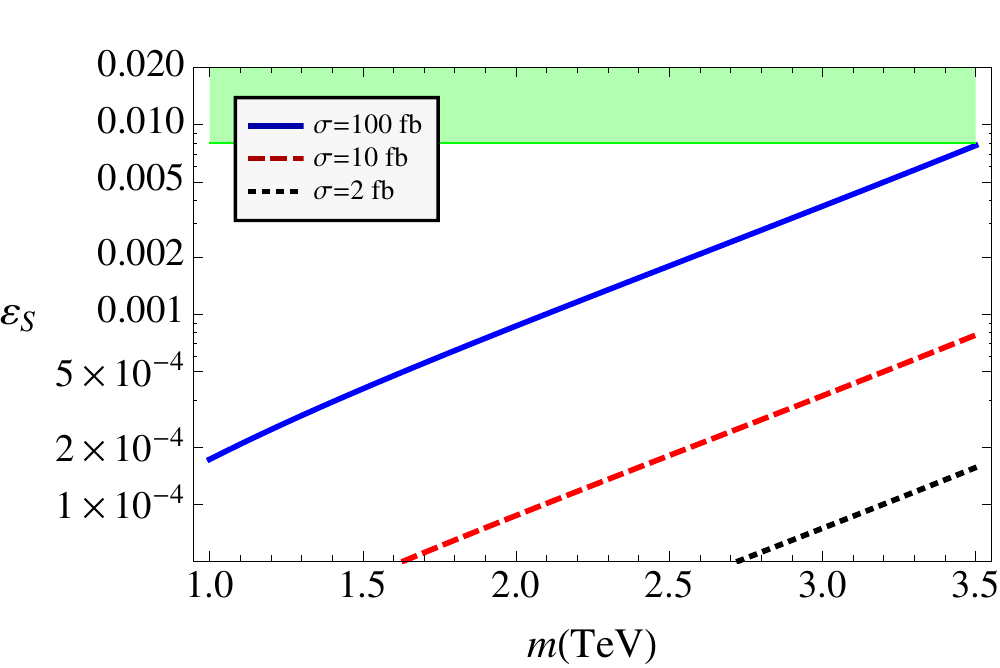}
\end{center}
\caption{Lower bounds on \(\epsilon_S\) at 2 GeV given collider discovery
  cross-sections of 100, 10, and 2 fb at a center-of-mass energy of
  \(\sqrt{s}=14\ {\rm TeV}\).}
\label{Collider}
\end{minipage}
\hfill\vrule width0pt
\end{figure}

The BSM couplings can be directly probed at colliders as excess large
transverse mass events in the channel \(p p \to e\bar\nu + X\). Using
the CMS report that the excess in this channel at \(m_T > 1\ {\rm
  TeV}\) is less than 3.7 events in \(1.13\ {\rm fb}^{-1}\) of data at
\(\sqrt s = 7\ {\rm TeV}\)~\cite{ref122}, we can, therefore, obtain
bounds on the BSM couplings.  As discussed later, and shown in
Fig.~\ref{BSMlow}, these bounds are currently weaker than the bounds
obtained from low energy experiments.

The collider bounds, however, get considerably stronger if the scalar
interaction is due to a resonance that is accessible at the LHC
energies.  If this new resonance couples to the quarks with a coupling
constant \(g_q\) and to the leptons with \(g_l\), then the partial
cross section can be written as
\begin{equation}
\sigma = g_q^2 g_l^2 \frac {m_S} {48 s\,\Gamma_S} L(\tau)\,,
\end{equation}
where \(s\) is the square of the center-of-mass energy, \(\tau \equiv
m^2/s\), \(L(\tau)\) is the relevant parton-distribution function and
\(m_S\) and \(\Gamma_S\) are the mass and width of the resonance.
With these couplings, the resonance can decay at least to quarks
and leptons, so we have
\begin{equation}
\Gamma_S \geq (g_l^2 + 2 N_c g_q^2) \frac m {16 \pi}\,,
\end{equation}
where \(N_c\) is the number of colors in the theory.  At low energies, the same couplings give a contribution to
\(\epsilon_S\):
\begin{equation}
\epsilon_S = 2 g_q g_l \frac{v^2}{m^2} \geq \frac {12 v^2 \sqrt{2
    N_c}} {\pi \tau L(\tau)} \sigma\,.
\end{equation}
As shown in Fig.~\ref{Collider}, this implies that for reasonable
discovery cross-sections of 100 fb at \(\sqrt s = 14\ {\rm TeV}\), a
low-energy measurement sensitivity of \(10^{-4}\) on \(\epsilon_S\) is
highly competitive.

\section{Neutron beta decay}
All of the BSM four-fermion operators contribute to the neutron beta
decay \(n (p_n) \to\allowbreak p (p_p)\allowbreak e^- (p_e)\allowbreak \bar{\nu}_e (p_\nu)\).  The
transition matrix elements of the quark bilinears required to analyze
this can be parameterized as~\cite{Weinberg:1958ut}
\begin{subequations}
\begin{eqnarray}
\br{p (p_p) } \bar{u} \gamma_\mu d \ket{n (p_n)} &=&
\bar{u}_p (p_p)  \left[
g_V(q^2)    \gamma_\mu    
+ \frac{\tilde{g}_{T(V)} (q^2)}{2 M_N}    \sigma_{\mu \nu}   q^\nu  
+ \frac{\tilde{g}_{S} (q^2)}{2 M_N}     q_\mu  
\right]  
 u_n (p_n)  \,,\nonumber\\&&
\\
\br{p (p_p) } \bar{u} \gamma_\mu \gamma_5  d \ket{n (p_n)} &=&
\bar{u}_p (p_p)  \left[
g_A(q^2)    \gamma_\mu    
+ \frac{\tilde{g}_{T(A)} (q^2)}{2 M_N}   \sigma_{\mu \nu}   q^\nu  
+ 
\frac{\tilde{g}_{P} (q^2)}{2 M_N}   q_\mu  
\right]  \gamma_5  u_n (p_n)  \,,\nonumber\\&&
\\
\br{p (p_p) } \bar{u}    d \ket{n (p_n)} &=&
g_S(q^2)  \bar{u}_p (p_p)   u_n (p_n) \,, 
\\
\br{p (p_p) } \bar{u}   \gamma_5   d \ket{n (p_n)} &=&
g_P(q^2)  \bar{u}_p (p_p)   \gamma_5  u_n (p_n) \,, 
\\
\br{p (p_p) } \bar{u}  \sigma_{\mu \nu}    d \ket{n (p_n)} &=&
 \bar{u}_p (p_p)   
 \left[
g_T(q^2)  \sigma_{\mu \nu}   +  g_{T}^{(1)} (q^2)  \left(q_\mu \gamma_\nu - q_\nu \gamma_\mu \right) 
\nonumber  \right. \\
&+& 
\left.  g_{T}^{(2)} (q^2)  \left( q_\mu P_\nu - q_\nu P_\mu  \right) 
+
g_{T}^{(3)} (q^2) 
 \left(
\gamma_\mu  \slashed{q}  \gamma_\nu - 
\gamma_\nu  \slashed{q} \gamma_\mu
 \right) 
\right] 
u_n (p_n)\,, 
\end{eqnarray}
\end{subequations}
where $u_{p,n}$ are the proton and neutron spinor amplitudes, $P = p_n
+ p_p$, $q = p_n - p_p$ is the momentum transfer, and $M_N= (M_n +
M_p)/2$ denotes an isospin-invariant nucleon mass.

Note that we are interested in disentangling the effects of
\(\epsilon_{P,S,T}\) which are expected to be about \(10^{-3}\) when
induced by BSM physics at the TeV scale.  This is the same size as the
recoil corrections of order \(q/M_N\), as well as the radiative
corrections proportional to \(\alpha_s/\pi\) and isospin breaking
effects proportional to \((M_n-M_p)/M_N\). In the above equation, all
the spinor contractions are \(O(1)\), except for \(\bar{u}_p \gamma_5
u_n\) which is $O(q/M_N)$. Furthermore, only the vector and axial
vector bilinears appear in the standard model, the rest are pure BSM
corrections and appear multiplied by \(\epsilon_{S,P,T}\). Finally,
the change in the form factors between zero momentum and the finite
recoil are proportional to \(q^2/\Lambda_{\rm QCD}^2 \sim
10^{-5}\). In light of this, we now discuss the contributions from
these bilinears that are relevant to the linear order in a
simultaneous expansion in \(\epsilon\), \(q/M_N\), \((M_n-M_p)/M_N\),
\(q^2/\Lambda_{\rm QCD}^2\) and \(\alpha_s/\pi\).
\begin{itemize}
\item{\bf Vector Current:} The form factor \(g_V(0)\) contributes to
  the leading order, whereas the weak magnetic charge
  \(\tilde{g}_{T(V)}(0)\) contributes to the first order in \(q/M_N\).
  The former is \(1\) up to second-order corrections in isospin
  breaking and the latter can be related to the difference of proton
  and neutron magnetic moments by isospin symmetry. Both of these are,
  therefore, known to the required accuracy. The induced-scalar form
  factor, \(\tilde g_S(0)\), vanishes in the isospin limit and is
  further proportional to \(q_\mu/M_N\), so it can be neglected to
  this order.
\item{\bf Axial Current:} \(g_A(0)\) contributes to this marix element
  at our required order.  The induced-tensor form factor,
  \(\tilde{g}_{T(A)}(0)\), vanishes in the isospin limit and has an
  explicit \(q_\mu/M_N\), whereas the induced pseudoscalar, \(\tilde
  g_P(0)\), is proportional both to \(q_\mu/m_N\) and to the
  pseudoscalar spinor contraction that it is also of order \(q_\mu/m_N\).
\item{\bf Pseudoscalar bilinear}: This entire term is subleading since
  the pseudoscalar contraction is proportional to \(q_\mu/m_N\) and
  the contribution is also proportional to a BSM coupling.
\item{\bf Scalar and Tensor bilinears}: The terms proportional to
  \(g_S(0)\) and \(g_T(0)\) are \(O(1)\) and multiplied by the BSM
  couplings \(\epsilon_S\) and \(\epsilon_T\).  The
  \(g_T^{(1,2,3)}\) contributions are subleading since they are
  multiplied by an explicit factor of \(q^\mu/m_N\).
\end{itemize}
In summary, the only matrix elements that feed into the leading order
determination of the BSM coefficients, and are not directly constrained by
experiments to the required order, are \(g_A(0)\), \(g_S(0)\), and
\(g_T(0)\).\footnote{The effect of \(\tilde g_P(0)\) is, however, only
  slightly smaller. Using PCAC relations, one can show that this
  matrix element is proportional to \(M_N/m_q \sim 100\). Its
  contribution to the amplitude is, therefore, about \(10^{-4}\)
  instead of the expected \(10^{-6}\).}  Furthermore, the BSM
coefficient \(\epsilon_V\) can be absorbed into a redefined Fermi
constant \(\tilde G_F^{(0)} \equiv G_F^{(0)} (1 + 2 \epsilon_V)\), and
\(\epsilon_A\) can similarly be used to redefine the ratio of axial
and vector charges: \(\tilde\lambda \equiv (\epsilon_V - \epsilon_A)
g_A(0)/g_V(0)\).

The differential decay distribution of the neutron is given
by~\cite{Ando:2004rk,Gudkov:2005bu}
\begin{eqnarray}
\frac{d\Gamma}{dE_e d \Omega_e d \Omega_\nu} &=& \frac{(\tilde G_F^{(0)})^2  \,  |V_{ud}|^2 }{(2\pi )^5}  
 \left( 1 + 3 \, \tilde{\lambda}^2 \right) 
\cdot  w (E_e) \cdot D  (E_e,  \mathbf{p} _e, \mathbf{p}_\nu, \boldsymbol{\sigma}_n)   \, ,   
\end{eqnarray}
where $ \mathbf{p} _e$ and $\mathbf{p} _\nu$ denote the electron and
neutrino three-momenta, and $\boldsymbol{\sigma}_n$ denotes the
neutron polarization.  The bulk of the electron spectrum is described
by
\[
w (E_e)  =  p_e E_e (E_0 - E_e)^2  \times {\rm rad.corr.}\,,
\]
where $E_0 = \Delta - (\Delta^2 - m_e^2)/(2 M_n)$ (with $\Delta = M_n
- M_p$) is the electron endpoint energy, $m_e$ is the electron mass,
and $\rm rad.corr.$ stands for the Coulomb and radiative
corrections~\cite{Ando:2004rk,Gudkov:2005bu,Czarnecki:2004cw}.
The remaining differential decay distribution function $D(E_e,
\mathbf{p} _e, \mathbf{p} _\nu, \boldsymbol{\sigma}_n)$ is parameterized
as~\cite{Ando:2004rk,Gudkov:2005bu,Gardner:2000nk}
\begin{equation}
D (E_e, \mathbf{p}_e, \mathbf{p} _\nu, \boldsymbol{\sigma}_n)    =   
1 +  c_0  + c_1 \frac{E_e}{M_N}   + \frac{m_e}{E_e} \bar{b}  +
\bar{B} (E_e)   \frac{\boldsymbol{\sigma}_n \cdot \mathbf{p
  }_\nu}{E_\nu} + \bar{A} (E_e)    \frac{\boldsymbol{\sigma}_n \cdot  \mathbf{p} _e}{E_e} 
+ \ldots\,,
\end{equation}
where \(c_{0,1}\) are recoil corrections, \(\bar b\) is a Fierz
interference term, \(\bar A(E_e)\) and \(\bar B(E_e)\) describe the
angular correlations between outgoing momenta and the neutron spin,
and the correlations between the outgoing electron and neutrino
momenta are not shown.\footnote{Recoil corrections to the asymmetry
  itself is discussed in Ref.~\cite{Sjue:2005ks}.}  An important point
to note is that experiments usually measure the angular dependence by
measuring the decay asymmetry, i.e. the decay rate in some `forward'
and `backward' bins normalized by the total decay rate.  Since the
Fierz intereference term appears in this normalization, extraction of
BSM contributions to these asymetries is always contaminated by the
BSM contributions to \(b\).

\begin{figure}[t]
\hfill
\includegraphics[height=0.2\vsize]{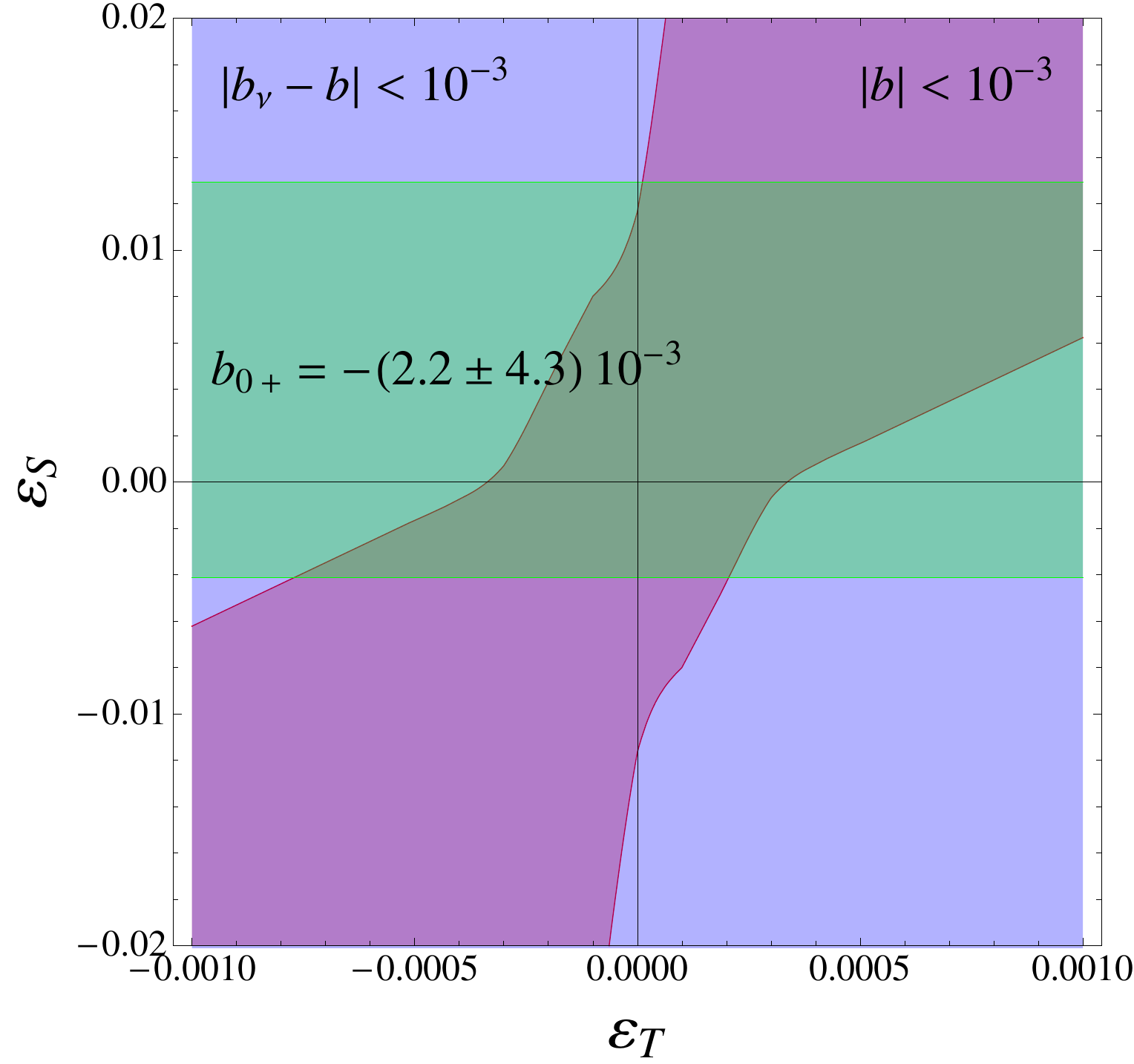}\hfill
\includegraphics[height=0.2\vsize]{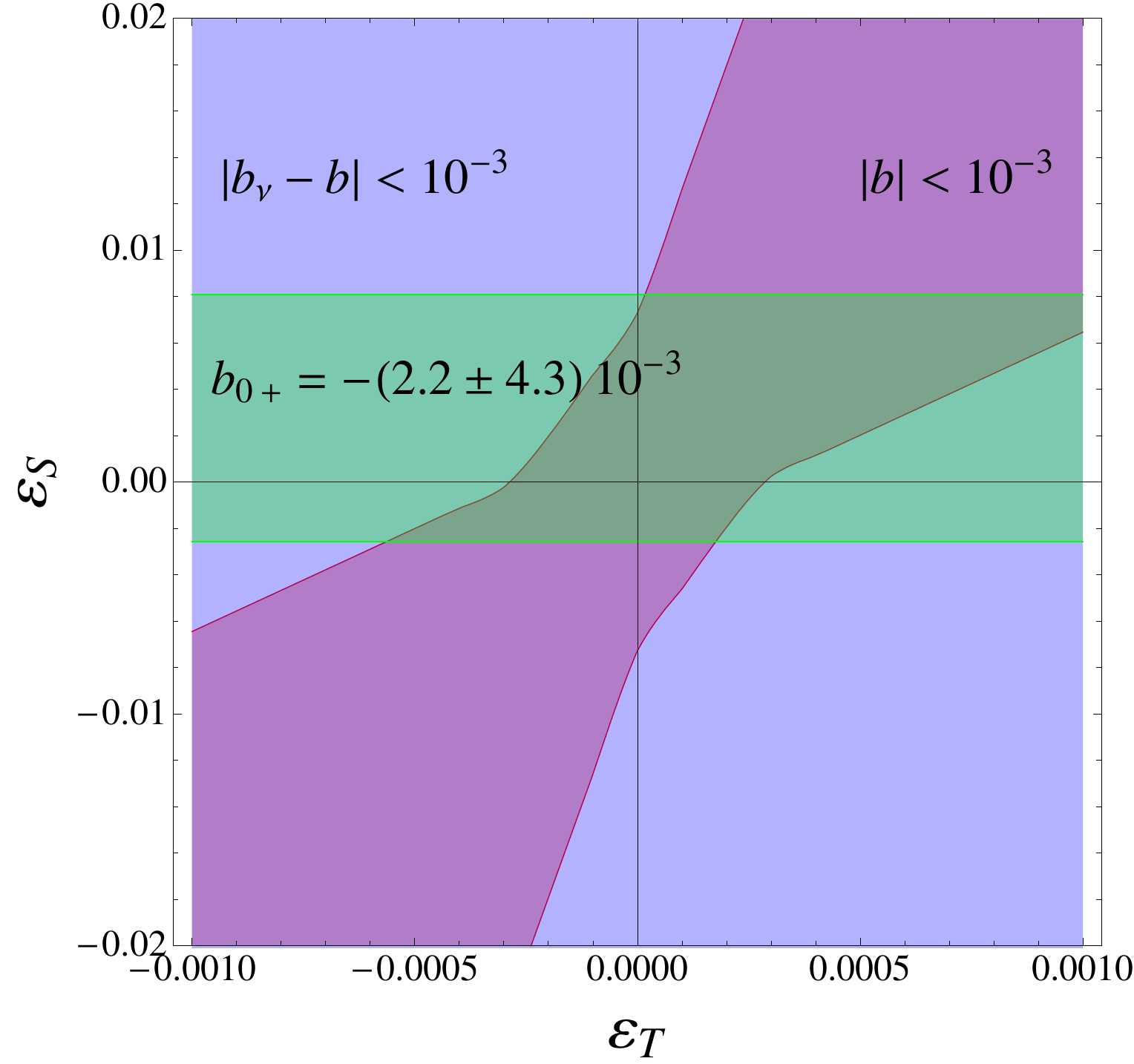}
\hfill\vrule width0pt
\caption{90\% Confidence Intervals of allowed regions in the
  \(\epsilon_S\)-\(\epsilon_T\) plane by the existing bounds on
  \(0^+\to0^+\) nuclear beta decay life times and the projected
  measurements of the neutron decay asymmetry at the \(10^{-3}\)
  level. The left panel shows the results with the scalar and tensor
  charges constrained by phenomenology, whereas the right panel uses
  the current lattice estimates.}
\label{fig8}
\end{figure}

The BSM scalar and tensor interactions appear to linear order in the
above decay matrix element in only two terms~\cite{ourpaper}:
\begin{subequations}
\begin{eqnarray}
  {\bar b}^{\rm BSM} & \approx & 0.34 g_S\epsilon_S - 5.22 g_T
  \epsilon_T\,,\\[10pt]
  {\bar b}_\nu \equiv \left.E_e \frac{\partial \bar{B}^{\rm BSM}(E_e)}{\partial m_e}\right|_{m_e=0}
  &\approx& 0.44 g_S \epsilon_S - 4.85 g_T \epsilon_T\,,
\end{eqnarray}
\end{subequations}
where all the matrix elements are evaluated at zero momentum
transfer. Currently both of these quantities have extremely weak
bounds: they are known to lie in the interval \([-0.3,0.5]\) at 95\%
Confidence Level.  Experiments to measure these quantities to the
level of \(10^{-3}\) are under way~\cite{RajanContrib}.  Additionally,
the scalar and tensor charges of the nucleon are very poorly
constrained by phenomenology~\cite{ref18}: \(0.25 < g_S < 1.0\) and
\(0.6 < g_T < 2.3\), and current lattice estimates also have large
uncertainties: \(g_S = 0.8(4)\) and \(g_T =
1.05(35)\)~\cite{ourpaper}.  In Fig.~\ref{fig8}, we show the impact of
a \(10^{-3}\) level measurement with these uncertainties on the
estimates of the charges, though lattice calculations are under way to
improves these estimates~\cite{HueyWenContrib}.

\section{Low Energy Phenomenology}
The BSM coefficients \(\epsilon_{S,P,T}\) are also constrained by
various nuclear beta decays.  In particular, the half lives of various
\(0^+ \to 0^+\) decays constrain the scalar coupling~\cite{HardyTowner}
to a 90\% Confidence Interval (CI) of \(-1 \times 10^{-3} < g_S
\epsilon_S < 3.2 \times 10^{-3}\).  The tensor coupling, on the other
hand, can be constrained by studying pure Gamow-Teller transitions;
the 90\% CI from $^{60}$Co and $^{114}$In are \(-2.9 \times 10^{-3} <
g_T\epsilon_T < 1.5 \times 10^{-2}\)~\cite{ref45} and
\(-2.2\times10^{-3} < g_T\epsilon_T < 1.3\times10^{-2}\)~\cite{ref46},
respectively.  Further bounds can be obtained by studying the angular
and momentum correlations in various beta decays, and some of the best
90\% CIs from such measurement are \(-0.76 \times 10^{-2} <
g_S\epsilon_S + 0.18 g_T\epsilon_T < 1.0 \times 10^{-2}\) and
\(|g_T\epsilon_T| < 3.1 \times 10^{-3}\) from positron polarization
measurements~\cite{ref47,ref48,ref49}, and \(| g_T\epsilon_T | < 6
\times 10^{-3}\) from beta-neutrino momentum
correlations~\cite{ref50}.

The pion decays are very precisely measured and can also be used to
constrain the BSM couplings.  In particular, the branching ratio of
pion decays to electrons,
\begin{equation}
R_\pi \equiv \frac{\Gamma(\pi \to e\nu[\gamma])}{\Gamma(\pi \to
  \mu\nu[\gamma])}\,,
\end{equation}
is very well constrained.  This BSM contribution is given as
\begin{equation}
\frac{R_\pi}{R_\pi^{SM}} = \frac{ \left( 1 - \frac
    B{m_e}\epsilon_P^{ee}\right)^2 + \left( \frac
    B{m_e}\epsilon_P^{e\mu}\right)^2 + \left( \frac
    B{m_e}\epsilon_P^{e\tau}\right)^2 }
{ \left( 1 - \frac
    B{m_\mu}\epsilon_P^{\mu\mu}\right)^2 + \left( \frac
    B{m_\mu}\epsilon_P^{\mu e}\right)^2 + \left( \frac
    B{m_\mu}\epsilon_P^{\mu\tau}\right)^2 }\,.
\end{equation}
Unless there are accidental cancellations, the quadratic terms in the
denominator can be neglected.  The contribution of the quadratic terms
in the numerator is, however, enhanced by the large coefficient
\(B/m_e \approx 3.6 \times 10^3\) in $\overline{\rm MS}$ at 1 GeV. The
experimental constraint \(R_\pi/R_\pi^{SM} = 0.996 \pm 0.005\) at 90\%
confidence then allows only a small spherical shell in
\(\epsilon_P^{ee,\mu,\tau}\) space that is centered at \(2.75 \times
10^{-4},0,0\) with a radius of \(2.75 \times 10^{-4}\) and a thickness
of \(1.38 \times 10^{-6}\).  Therefore, without assuming any relation
between the various pseudoscalar couplings, one can only bound them as \looseness-1
\begin{subequations}
\begin{equation}
-1.4 \times 10^{-7} {}< \epsilon_P^{ee} <{} 5.5 \times 10^{-4}\,,\qquad
-2.75 \times 10^{-4} {}< \epsilon_P^{e\mu,\tau} <{} 2.75 \times 10^{-4}\,.
\end{equation}
\end{subequations}
Standard model radiative corrections, however, mix the scalar, tensor,
and pseudoscalar couplings:
\begin{eqnarray}
\epsilon_P(\mu) &=& \epsilon_P(\Lambda) \left(1 + 1.3 \times 10^{-2}
  \log\frac\Lambda\mu\right)
  \nonumber\\  &&\quad
+ 6.7\times 10^{-4}\epsilon_S(\Lambda) \log\frac\Lambda\mu
     - 7.3 \times 10^{-2}\epsilon_T(\Lambda) \log\frac\Lambda\mu\,,
\end{eqnarray}
where we have suppressed the family indices. As a result, barring
cancellations, the stringent constraints on the pseudoscalar coupling
translate to constraints on scalar and tensor couplings as well:
\(|\epsilon_S| \lesssim 8 \times 10^{-2}\) and \(|\epsilon_T| \lesssim
10^{-3}\).  The constriaint on the tensor is similar to that obtained
directly from the radiative branching fraction of the pion decay: \(-2
\times 10^{-4} < \epsilon_T f_T < 2.6 \times 10^{-4}\), where \(f_T\),
the tensor charge of the pion, is estimated to be \(0.24\pm0.04\).

\section*{Acknowledgements}
We thank V.~Cirigliano, A.~Filipuzzi, M.~Gonzalez-Alonso and
M.~Graesser, who collaborated with us on the detailed paper~\cite{ourpaper}.  The
speaker is supported by the DOE grant DE-KA-1401020.

\end{document}